\documentclass[trackchanges,twocolumn,groupedaddress]{aastex701}

\usepackage{amsmath,bm,hyperref}
\usepackage{cleveref} 
\usepackage{wasysym}

\crefrangelabelformat{equation}{(#3#1#4--#5#2#6)}

\begin{document}

\title{Constraining Planet Nine with Pulsar Timing: Orbit Determination Method from the Acceleration, Jerk, and Snap of the Solar System Barycenter}
\shorttitle{Constraining Planet Nine with Pulsar Timing}
\shortauthors{Taylor, Fonseca, \& Glaser}

\author[0009-0006-8984-9220]{Jackson D. Taylor}
\affiliation{Department of Physics and Astronomy, West Virginia University, P.O. Box 6315,
Morgantown, WV 26506, USA}
\affiliation{Center for Gravitational Waves and Cosmology, West Virginia University, Chestnut Ridge Research Building, Morgantown, WV 26505, USA}
\email{jdt00012@mix.wvu.edu, jacksondtaylor4@gmail.com}

\author[0000-0001-8384-5049]{Emmanuel Fonseca}
\affiliation{Department of Physics and Astronomy, West Virginia University, P.O. Box 6315,
Morgantown, WV 26506, USA}
\affiliation{Center for Gravitational Waves and Cosmology, West Virginia University, Chestnut Ridge Research Building, Morgantown, WV 26505, USA}
\email{emmanuel.fonseca@mail.wvu.edu}

\author[0000-0003-4090-9780]{Joseph Glaser}
\affiliation{Department of Physics and Astronomy, West Virginia University, P.O. Box 6315, Morgantown, WV 26506, USA}
\affiliation{Center for Gravitational Waves and Cosmology, West Virginia University, Chestnut Ridge Research Building, Morgantown, WV 26505, USA}
\email{joseph.glaser@mail.wvu.edu}

\correspondingauthor{Jackson D. Taylor}
\email{jdt00012@mix.wvu.edu, jacksondtaylor4@gmail.com}

\begin{abstract}


Planet Nine is hypothesized to exist in the outer solar system in order to explain the physical orbital clustering of several trans-Neptunian objects. A Planet Nine would systematically change the location of the solar system barycenter (SSB). 
Therefore, the precise timing properties of millisecond radio pulsars would measure the acceleration, jerk, and snap of the currently-perceived SSB under the gravitational influence of a Planet Nine. 
We present a method on how to use pulsar spin frequency derivatives and binary pulsar orbital period derivatives to constrain the Planet Nine parameter space.
We find that searches for unidentified massive objects in the solar system require at least cubic and quartic variations in the pulse times of arrival (corresponding to jerk and snap, respectively) in order to uniquely constrain an orbit. 
Otherwise, additional assumptions are required to fully constrain the mass and orbital parameters of the object. 
We use our method to weigh the ice giants Uranus and Neptune and obtain results consistent with the literature.
This method could be used to weigh a Planet Nine after a putative optical detection. 
Future work will use updated pulsar timing datasets that benefit from longer observational baselines and advances in statistical modeling and noise characterization.

\end{abstract}


\keywords{\uat{Solar system planets}{1260} --- \uat{Pulsar timing method}{1305} --- \uat{Millisecond pulsars}{1062} --- \uat{Celestial mechanics}{211} --- \uat{Planetary dynamics}{2173}}

\section{Introduction}

The solar system census is in a period of rapid growth.
Wide-field searches continue to add distant bodies on extreme orbits, including several trans-Neptunian objects (TNOs), such as Sedna-like objects (SNOs) and \mbox{(Neptune-)detached} Kuiper Belt Objects (KBOs; e.g., \citealt{trujillo_sheppard2014sedna, theplanetninehypothesisBB19}).
The Vera C. Rubin Observatory's Legacy Survey of Space and Time \citep{ivezic2019_lsst_vera_rubin} will expand this census over the coming decade. Each such discovery sharpens our picture of the outer solar system and is a reminder that the mass inventory defining the solar system barycenter (SSB) is incomplete.

One of the most interesting aspects of this growing population of distant celestial bodies is that the most-distant members appear to be clustered in orbital axis (i.e., their apsidal lines) and angular momentum vector \citep{trujillo_sheppard2014sedna, batygin2016evidenceBB16}. Further, \citet{batygin2016evidenceBB16} established the modern Planet Nine (P9) hypothesis by showing that an eccentric, inclined planet with mass $m_9\sim10$M$_\oplus$ could explain the detached KBO apsidal and orbital plane clustering and confinement.
Further analysis that explicitly models survey selection placed the probability that a uniform population would appear this clustered at $0.2\%$ \citep{brown2019orbital_BB19_clustering}.
This probability was later revised to $0.4\%$, together with a reference population of P9 orbits that explain the known TNOs to varying probabilities (\citealt{brown2021orbit_BB21}; BB21 hereafter).
Restricting attention to dynamically stable TNOs, the more recent \citet{siraj2025orbit} found ${\sim}3\sigma$ clustering in the longitude of perihelion alone, and prefer a planet of mass $4.4\pm1.1\,\mathrm{M}_\oplus$ at semi-major axis $290\pm30$\,AU that lies almost entirely outside that earlier reference population. Thus, the P9 mass and distance are unsettled, and likely will remain so until a positive optical detection.

The evidence against the clustering hypothesis is equally persistent. The Outer Solar System Origins Survey, whose detection biases were characterized by design, found its own distant TNOs consistent with uniformly distributed orbital angles \citep{shankman2017ossos}.
\citet{napier2021no} also find no evidence for angular clustering after analyzing the biases of three surveys.
Recent discoveries add to the tension. The TNO 2017~OF$_{201}$ sits outside the apsidal cluster and is ejected in a majority of simulations containing the BB21 P9 \citep{cheng2026discovery_2017_OF201}.
Further, 2023~KQ$_{14}$ is misaligned with the other SNOs and pushes any surviving P9 to a more distant orbit \citep{chen2025discovery}. Explanations requiring no planet at all, such as a massive self-gravitating outer belt \citep{zderic2020giant} or a stellar flyby in the young solar system \citep[e.g.,][]{pfalzner2026TNOs_via_stellar_flyby}, remain viable. 
After a decade, the debate has largely stalled due to the still relatively small number of known detached KBOs. 

Measurements of the SSB through pulsar timing offer an orthogonal line of evidence. Pulsars are rapidly rotating neutron stars with beamed emission out of their off-rotation-axis magnetic poles. 
By insisting that each observed pulse of emission marks exactly an integer number of rotations since the last observed pulse, pulse times of arrival (TOAs) have turned pulsars into high-precision clocks \citep[e.g.,][]{APTBtaylor2024}, a process known as pulsar timing. Millisecond pulsars (MSPs) in particular can measure variations down to the $\mathcal{O}(100\,\mathrm{ns})$ scale \citep[e.g.,][]{agazie2023nanograv}.

Using pulsars as accelerometers and, in general, dynamical probes is not new. Pulsars in binaries with other stars (known as binary pulsars) have famously been able to measure the mass of neutron stars \citep{cromartie2020relativistic} and confirm General Relativity to the highest precisions \citep[e.g.,][]{kramer2021strong}. 
Measuring the changes in the binary orbital period of binary MSPs has made possible constraints on the Galactic potential \citep{donlanv3}, on pHz continuous gravitational waves \citep{DeRocco_pHz, zheng2026_pHz}, and even on candidate dark matter sub-halos in the Galaxy \citep{chakrabarti2026constraints_DM_subhalo}. On a per-pulsar basis, pulsars have probed the dynamics induced by exoplanets \citep{wolszczan1992_firstexoplanet, behrens_nanograv_2020}, gravitational scattering with interstellar freely-floating planets \citep{ISO_Ross_2020, ISO_loki_2026}, and co-orbital exotrojans \citep{Trojans_Jackson_2026}. 


TOAs are placed in the frame of the SSB, the location of which relies on accurately measured masses and positions of the largest solar system bodies. Therefore, an error in any object's mass or position would induce a sky-dipolar pattern in TOA residuals. 
This fact allowed \citet{caballero2018studying} (C+18 hereafter) to constrain the masses of the planets and a handful of the minor planets and \citet{guo2018dynamical_GTC18} to outline a method to search for unidentified massive objects (UMOs).
In contrast to TOA signals from UMOs, gravitational waves are expected to induce quadrupolar TOA variations \citep{hellings_downs_1983}. 
Further, planetary ephemerides fit to spacecraft ranging can only probe a P9 through its tidal field ($\propto m_9/r^3$; \citealt{fienga2016constraints,fienga2020new}), whereas pulsars, as external references, are sensitive to the uniform acceleration of the SSB itself ($\propto m_9/r^2$) and its time derivatives ($\propto m_9/r^{7/2}$, $\propto m_9/r^{5}$, $\propto m_9/r^{13/2}$, and so on; see Section~\ref{sec:methods}).

Given pulsar timing baselines of no more than 60\,yr currently, TOAs have not captured the full expected several-thousand-year orbital period of a P9. Thus, the first hints of a P9 in TOA datasets will not be revealed by the full sky-dependent waveforms, but rather through a Taylor series in the P9-induced motion of the SSB, measured through the Doppler terms of pulsar spin frequencies $(f_s)$ and binary orbital periods $(P_b)$. 
For clarity, we will refer to the currently-known solar system barycenter as the SSB, and the SSB-P9 joint barycenter as the SSP9B (solar system Planet Nine barycenter).
The time derivatives $\dot{f}_s$ and $\dot{P}_b$ correspond to the acceleration of the SSB, $\ddot{f}_s$ corresponds to its jerk, $\dddot{f}_s$ corresponds to its snap, and so on. 
\citet{Zakamska_2005} first constrained the pulsar-measured SSB acceleration from a distance orbiter using $\dot{f}_s$ and $\dot{P}_b$ measurements.
Using Doppler terms to constrain long-period binary MSP orbits was later proposed and used by \citet{bassa2016millisecond,kaplan2016psr}.
Notably, we ignore $\dot{f}_s$ in this study as a P9-induced $\dot{f}_s$ is highly degenerate with pulsar spindown due to magnetic braking, a process that is intrinsic to each pulsar. 
\citet{wang2026_Pbdot_recent} find that $\dot{P}_b$ gives more precise SSB acceleration measurements than those from quasars (Gaia Collaboration et al. \citeyear{gaia2021_quasars}).
Further note that spin Doppler terms do not require that the pulsar be in a binary.

Thus, for this initial study, we present a method of constraining the P9 parameter space that does not use TOAs directly, but instead uses $\dot{P}_b$, $\ddot{f}_s$, and $\dddot{f}_s$ measurements. 
Unlike previous work, we explore how many Doppler terms are needed to fully constrain a P9. Our method uses data that easily flows from pulsar measurements derived for other purposes. Furthermore, the Taylor series model allows for analytical formulations of covariances and degeneracies arising from low signal-to-noise and short timing baseline data.


We develop our arguments in full below. In Section~\ref{sec:methods} we derive $\bm{a}_\odot$, $\bm{j}_\odot$, and $\bm{s}_\odot$ as functions of the P9 orbital elements and then invert that relation, showing explicitly that the three position derivatives uniquely determine the orbit. 
Section~\ref{sec:accel_jerk_degen} characterizes the one-parameter families of the P9 space when snap measurements are omitted, and shows that fixing the eccentricity, an apparently natural choice, does not uniquely specify one orbit. 
Section~\ref{sec:estimated_constraint_space} estimates what portions of the P9 space currently compiled datasets can constrain.
Section~\ref{sec:uranus_neptune} outlines how to weigh an optically detected P9 by weighing Uranus and Neptune.
We conclude with a discussion in Section~\ref{sec:discussion_and_conclusions}.

\section{Methods}
\label{sec:methods}

The orbital period $P_b$ of a binary pulsar may change with time due to changes in the orbital energy \citep{Damour_Taylor_1991}, changes in the viewing angle, or kinematic relative accelerations under the Galactic potential or a P9.
The observed variation can be written as
\begin{equation}
    \label{eq:pbdot_contributions}
    \dot{P}_b^{\rm Obs} = \dot{P}_b^{\rm GR} + \dot{P}_b^{\rm Shk} + \dot{P}_b^{\rm MW} + \dot{P}_b^{\rm 9},
\end{equation}
where $\dot{P}_b^{\rm GR}$ is the change in $P_b$ from general relativity \citep{Peters_1964}, $\dot{P}_b^{\rm Shk}$ is the Shklovskii term \citep{Shklovskii_1970}, $\dot{P}_b^{\rm MW}/P_b= -(\Delta \bm{a}_{\rm MW} \cdot \bm{\hat{n}})/c$ is due to the differential acceleration of the SSB and the pulsar under the Galactic potential. $\bm{\hat{n}}$ is the unit vector pointing from the SSB to the pulsar. $\dot{P}_b^{\rm 9}$ is due to the P9-induced acceleration on the SSB $\bm{a}_{\odot}$ via
\begin{equation}
    \label{eq:doppler_accel}
    -\frac{\dot{P}_b^{9}}{P_b} = \frac{\dot{f}_b^9}{f_b} = (\bm{a}_{\odot} \cdot \bm{\hat{n}})/c
\end{equation}

The spin frequency of a pulsar $f_s$ is also known to change with time, and if its second frequency derivative $\ddot{f}_s^{\rm Obs}$ is entirely attributed to a P9-induced jerk on the SSB $\bm{j}_{\odot}$, then
\begin{equation}
    \label{eq:doppler_jerk}
    \frac{\ddot{f}_s^{\rm Obs}}{f_s} = \frac{\ddot{f}_s^9}{f_s} \approx (\bm{j}_{\odot} \cdot \bm{\hat{n}})/c.
\end{equation}
The same follows for snap,
\begin{equation}
    \label{eq:doppler_snap}
    \frac{\dddot{f}_s^{\rm Obs}}{f_s} = \frac{\dddot{f}_s^9}{f_s} \approx (\bm{s}_\odot \cdot \bm{\hat{n}})/c,
\end{equation}
and so on for higher Doppler and position derivative terms. 
\Cref{eq:doppler_accel,eq:doppler_jerk,eq:doppler_snap} assume $v_9/c \ll 1$ where $v_9$ is the speed of P9 with respect to the SSB. 

A major advantage of the jerk observable ($\ddot{f}_s$), or ``channel", over the acceleration observable ($\dot{P}_b$) is that it is largely free of the contaminating terms in \Cref{eq:pbdot_contributions}. The Galactic jerk on the SSB is $|\bm{j}_{\rm MW}|/c \sim (|\bm{a}_{\rm MW}|/c)\,\Omega_{\rm MW} \sim 10^{-33}\,\mathrm{s^{-2}}$ for a ${\sim}230$\,Myr Galactic orbit, giving $\ddot{f}_s \sim 10^{-31}\,\mathrm{Hz^3}$ for a $f_s = 200$\,Hz pulsar. 
Intrinsic magnetic braking contributes $\ddot{f}_s \lesssim 10^{-32}\,\mathrm{Hz^3}$ for typical MSPs \citep{liu2018high}. 
Kinematic terms arising from the pulsar's radial velocity and proper motion are typically $\sim10^{-31}\,\mathrm{Hz^3}$ (\citealt{liu2019correlated}; L+19 hereafter), but reach $10^{-30}$ to $10^{-28}\,\mathrm{Hz^3}$ for nearby, high proper-motion pulsars (e.g., PSRs~J0437$-$4715 and J1909$-$3744; \citealt{liu2018high}), comparable to the $\ddot{f}_s \sim 10^{-30}$ to $10^{-27}\,\mathrm{Hz^3}$ expected across the P9 parameter space of interest (see Section~\ref{sec:estimated_constraint_space}). Further, pulsars in globular clusters \citep[e.g.,][]{Padmanabh_Terzan5_2024} should not be used for jerk measurements due to the strong local potential intrinsic to the cluster.
The dominant remaining limitation is red timing noise, which requires careful modeling (e.g., L+19).

The relevant parameters of P9 are its mass $m_9$, semi-major axis in the SSB frame $a_9$, longitude of ascending node $\Omega$, argument of perihelion $\omega$, inclination $i$, eccentricity $e$, and true anomaly $\nu$. Other parameterizations are also possible. Let $\bm{\theta} = \{a_9, m_9, \Omega, \omega,  i, e, \nu \}$, which is the relevant P9 parameter space accessible through pulsar timing. We will now show how the P9-induced acceleration, jerk, and snap on the SSB (i.e., the position derivatives $\bm{\mathcal{P}} \equiv \{ \bm{a}_{\odot}, \bm{j}_{\odot}, \bm{s}_{\odot} \}$) uniquely determine the orbit of P9.

\begin{figure*}
  \centering
  \includegraphics[width=.94\textwidth]{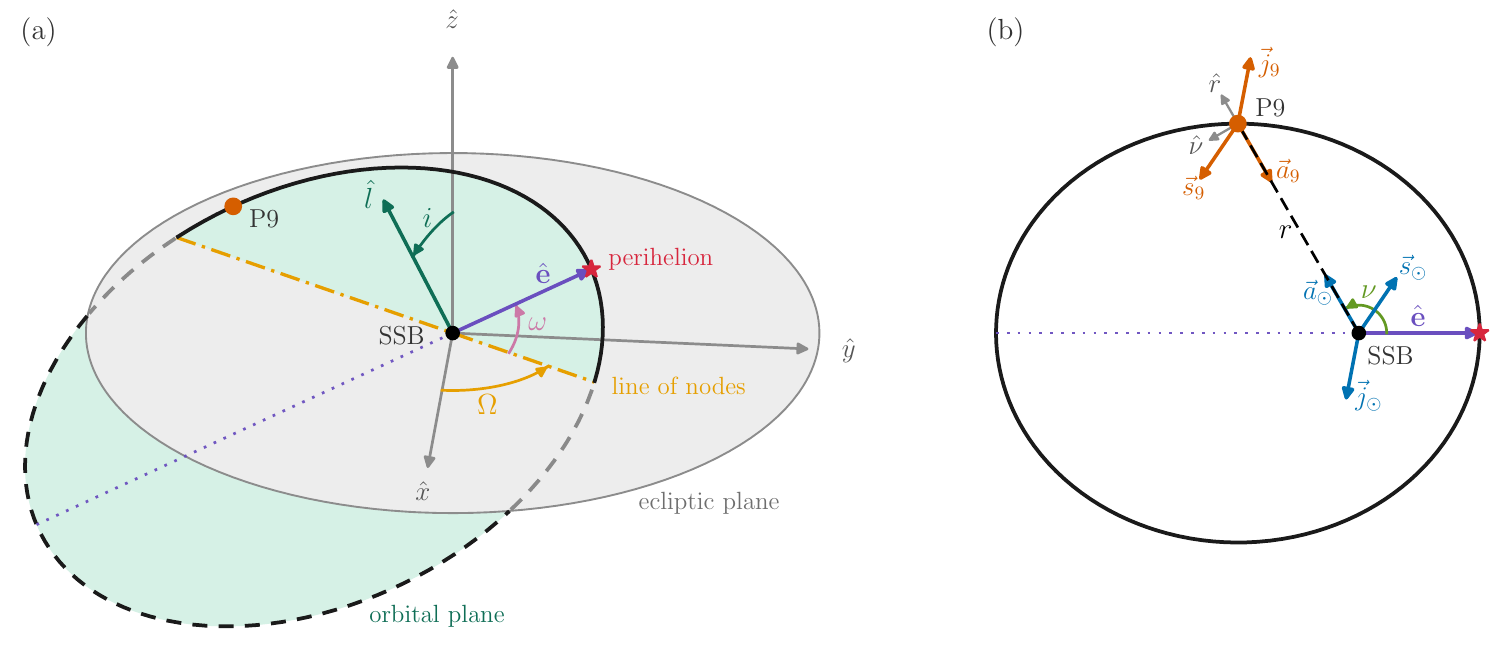}
  \caption{Geometry of the P9-SSB system, drawn for the representative configuration $e = 0.5$, $i = 30^\circ$, $\Omega = 60^\circ$, $\omega = 50^\circ$, and $\nu = 120^\circ$. 
  (a) The angles that define the P9 orbital plane with respect to the ecliptic plane. The unit vectors $\bm{\hat{z}}$ and $\bm{\hat{l}}$ are normal to the ecliptic and P9 orbital planes, respectively. The apsidal line is dotted and the line of nodes is dash-dotted. 
  (b) The relevant orbital-plane vectors. The direction of each position-derivative vector was determined exactly via \Cref{eq:accel_ssb,eq:jerk_ssb,eq:snap_ssb_elements} for the chosen configuration, with arbitrary lengths. The unit vector $\bm{\hat{l}}$ is directed out of the page, and therefore P9 is shown orbiting counter-clockwise. The SSB-P9 distance line is dashed.}
  \label{fig:geometry}
\end{figure*}

We point the reader to Figure~\ref{fig:geometry} to aid in the following discussions. 
Let $\hat{\bm{x}}$, $\hat{\bm{y}}$, and $\hat{\bm{z}}$ be the SSB ecliptic unit vectors, where $\hat{\bm{x}}$ points to the vernal equinox, $\hat{\bm{z}}$ points to the ecliptic north pole (defined by the Earth's orbital angular momentum vector), and $\hat{\bm{y}} = \hat{\bm{z}} \times \hat{\bm{x}}$ to complete the right-handed coordinate system. Let $\hat{\bm{e}}$, $\hat{\bm{Y}}$, and $\hat{\bm{l}}$ be the cartesian unit vectors defined in the P9-orbital plane as
\begin{align}
    \label{eq:coord_transf}
    \hat{\bm{e}} &= \bm{R}_{\Omega, i, \omega}\,\hat{\bm{x}}, \qquad\\
    \hat{\bm{Y}} &= \bm{R}_{\Omega, i, \omega}\,\hat{\bm{y}}, \qquad\\
    \hat{\bm{l}} &= \bm{R}_{\Omega, i, \omega}\,\hat{\bm{z}},
\end{align}
where 
\begin{equation}
    \label{eq:rotation}
    \bm{R}_{\Omega, i, \omega}
    =
    \begin{bmatrix}
        c_\omega c_\Omega - c_i s_\Omega s_\omega
        & -s_\omega c_\Omega - c_i s_\Omega c_\omega
        & s_i s_\Omega \\[4pt]
        c_\omega s_\Omega + c_i c_\Omega s_\omega
        & -s_\omega s_\Omega + c_i c_\Omega c_\omega
        & -s_i c_\Omega \\[4pt]
        s_i s_\omega
        & s_i c_\omega
        & c_i
    \end{bmatrix}.
\end{equation}
For compactness, we adopt the notation $c_x \equiv \cos x$ and $s_x \equiv \sin x$ for $x \in \{\Omega, i, \omega\}$.
Lastly, let $\bm{\hat{r}}$ and $\bm{\hat{\nu}}$ be the polar coordinates in the P9 orbital plane where $\bm{\hat{r}}$ points from SSB to P9 and $\bm{\hat{\nu}} = \bm{\hat{l}} \times \bm{\hat{r}}$ is the azimuthal direction. 

\subsection{Acceleration and Jerk}

The acceleration of P9 under the gravitational influence of the Sun and the eight major planets 
\begin{equation}
    \bm{a}_9 = -\sum_{k=0}^{8}\frac{Gm_k}{\lVert\bm{r}_9 - \bm{r}_k\rVert^3}(\bm{r}_9 - \bm{r}_k),
\end{equation}
where $m_k$ is the mass of object $k$ and $\bm{r}_k$ is the position of object $k$ with respect to the SSB, where $k=0$ is the Sun, $k=1$ is Mercury, ..., $k=8$ is Neptune, and $k=9$ is P9. We will use the monopolar approximation $r_{k \neq 9} \ll r_9$, meaning 
\begin{equation}
    \label{eq:accel_ssb}
    \bm{a}_\odot= -\frac{m_9}{\mathrm{M}_{\mathrm{SS}}}\bm{a}_9 = \frac{Gm_9}{r^2}\hat{\bm{r}},
\end{equation}
where $\mathrm{M_{SS}} = \sum _{k=0}^{8}m_k$ is the approximately the mass of the known solar system, $\bm{r} \equiv \bm{r}_9$, $r = \lVert\bm{r}\rVert$, and $\bm{\hat{r}} = \bm{r}/r$.
The radial coordinate $r$ is a function of $a_9$, $e$, and $\nu$ via 
\begin{equation}
    r = \frac{a_9(1-e^2)}{1+e\cos \nu}.
\end{equation}

With $\dot{\bm{\hat{r}}} = \dot{\nu} \hat{\bm{\nu}}$,
where $\hat{\bm{\nu}}$ is the azimuthal direction in the plane of P9's orbit, 
\begin{equation}
\dot{\nu} = \sqrt{\frac{GM}{a_9(1-e^2)}}\frac{1+e\cos\nu}{r}, 
\end{equation}
and 
\begin{equation}
    \dot{r} = \sqrt{\frac{GM}{a_9(1-e^2)}}\, e\sin\nu,
\end{equation}
we have
\begin{align}
    \nonumber
    \bm{j_\odot}  =& \frac{d}{dt}\bm{a}_\odot\\
    \nonumber
    =&\frac{(1+e\cos\nu)^3\sqrt{GM}}{(1-e^2)^{7/2}}\frac{Gm_9}{a_9^{7/2}} \\
    \label{eq:jerk_ssb}
    &\times\left[ (-2e\sin\nu) \hat{\bm{r}} + (1+e\cos\nu)\bm{\hat{\nu}}\right],
\end{align}
where $M=M_{\rm SS} + m_9$.
Note that we defined $\hat{\bm{\nu}} = \bm{\hat{l}} \times \bm{\hat{r}}$ to be in the direction of the P9's orbital velocity component perpendicular to $\hat{\bm{r}}$, $\bm{v}_{9,\perp}$. For a circular orbit \Cref{eq:jerk_ssb} reduces to $\bm{j}_\odot (e=0) = (\sqrt{GM}\,Gm_9/a_9^{7/2})\,\hat{\bm{\nu}}$ which runs antiparallel to the SSB perpendicular velocity around the SSP9B.

\subsection{Snap}

Equations~\eqref{eq:accel_ssb} and~\eqref{eq:jerk_ssb} give six equations for seven unknowns ($\bm{\theta}$), leaving a one-parameter family of solutions. A positively measured $\bm{s}_\odot$ breaks this degeneracy.

In deriving $\bm{s}_\odot$, we find it useful to write $\bm{j}_\odot$ as
\begin{equation}
    \label{eq:jerk_PQ}
    \bm{j}_\odot = A\left(P\,\hat{\bm{r}} + Q\,\hat{\bm{\nu}}\right),
\end{equation}
where
\begin{align}
    \label{eq:P}
    P &\equiv j_{r}/A = -\frac{2\dot{r}}{r}, \\
    \label{eq:Q}
    Q &\equiv j_{\nu}/A = \dot{\nu}.
\end{align}
Here, $A = \lVert\bm{a}_\odot\rVert$ and $j_{\xi} = \bm{j}_\odot \cdot \bm{\hat{\xi}}$ for $\bm{\hat{\xi}} \in \{\bm{\hat{r}}, \bm{\hat{\nu}} \}$.
Since $\dot{A} = AP$,
$\dot{\hat{\bm{r}}} = Q\,\hat{\bm{\nu}}$, and $\dot{\hat{\bm{\nu}}} = -Q\,\hat{\bm{r}}$,
\begin{align}
    \label{eq:snap_raw}
    \bm{s}_\odot &= \frac{d}{dt}\left(\bm{j}_\odot \right)\\
    &= A\left[\left(P^2 + \dot{P} - Q^2\right)\hat{\bm{r}}
    + \left(2PQ + \dot{Q}\right)\hat{\bm{\nu}}\right].
\end{align}
We find that $\dot{Q} = PQ$ and $\dot{P} = P^2/2 - 2Q^2+ 2GM/r^3$.
Collecting terms,
\begin{equation}
    \label{eq:snap_PQ}
    \bm{s}_\odot = \frac{Gm_9}{r^2}
    \left[\left(\tfrac{3}{2}P^2 - 3Q^2 + \frac{2GM}{r^3}\right)\hat{\bm{r}}
    + 3PQ\,\hat{\bm{\nu}}\right],
\end{equation}
or, in orbital elements,
\begin{align}
    \nonumber
    \bm{s}_\odot =& \frac{(1+e\cos\nu)^4\,GM}{(1-e^2)^5}\frac{Gm_9}{a_9^5} \\
    \label{eq:snap_ssb_elements}
    \nonumber
    &\times\Big\{\left[6e^2\sin^2\nu - 3(1+e\cos\nu)^2 + 2(1+e\cos\nu)\right]\hat{\bm{r}}\\
    &\phantom{\times\Big\{}- 6e\sin\nu\,(1+e\cos\nu)\,\hat{\bm{\nu}}\Big\}.
\end{align}
For a circular orbit Eq.~\eqref{eq:snap_ssb_elements} reduces
to $\bm{s}_\odot (e = 0) = -(Gm_9GM/a_9^5)\,\hat{\bm{r}} \propto - \bm{a}_\odot$.

\subsection{Solving for Orbital Elements Given the Position Derivatives}
\label{sec:inverting_P_of_theta}

\Cref{eq:accel_ssb,eq:jerk_ssb,eq:snap_ssb_elements} state how $\bm{\mathcal{P}}$ is a function of $\bm{\theta}$. 
We will now find $\bm{\theta}(\bm{\mathcal{P}})$ to prove that the inversion is unique.
The acceleration fixes $\hat{\bm{r}}$, and $\hat{\bm{l}}$ follows from the co-planarity of $\bm{a}_\odot$ and $\bm{j}_\odot$:
\begin{align}
    \label{eq:basis_r}
    \hat{\bm{r}} &= \frac{\bm{a}_\odot}{A}, \\
    \label{eq:basis_l}
    \hat{\bm{l}} &= \frac{\bm{a}_\odot \times \bm{j}_\odot}
                        {\lVert\bm{a}_\odot \times \bm{j}_\odot\rVert}.
\end{align}
The relation $\bm{\hat{\nu}} = \bm{\hat{l}} \times \bm{\hat{r}}$ follows immediately. 
\Cref{eq:jerk_PQ,eq:P,eq:Q,eq:snap_PQ} show that two components of $\bm{s}_\odot$ do not add information not already present in $\bm{a}_\odot$ and $\bm{j}_\odot$. Namely,
\begin{align}
    \label{eq:snap_null}
    s_l &= \bm{s}_\odot \cdot \hat{\bm{l}} = 0,\\
    s_\nu  &= \bm{s}_\odot \cdot \hat{\bm{\nu}} = \frac{3\,j_r j_\nu}{A}.
\end{align}
The radial component however retains an explicit dependence on $M/r^3$,
\begin{equation}
    \label{eq:snap_radial}
    s_r = \bm{s}_\odot \cdot \hat{\bm{r}} = \frac{3j_r^2 - 6j_\nu^2}{2A} + \frac{2GMA}{r^3},
\end{equation}
which allows us to determine $r$ as follows. Writing
$D \equiv s_r - \left(3j_r^2 - 6j_\nu^2\right)/2A$ and noting that $m_9 = Ar^2/G$,
\begin{equation}
    \label{eq:r_cubic}
    D\,r^3 - 2A^2 r^2 - 2A\,GM_{\rm SS} = 0.
\end{equation}
The quantities $D= 2GMA/r^3$, $2A^2$, and $2AGM_{\rm SS}$ are positive, so the above cubic has a single positive root. 

Therefore,
\begin{align}
    \label{eq:mass_from_r}
    m_9 &= \frac{Ar^2}{G},\\
    \label{eq:i_from_l}
    i &= \arccos{ \left(\hat{l}_z\right)},\\
    \label{eq:Omega_from_l}
    \Omega &= \mathrm{atan2}{ \left(\hat{l}_x, -\hat{l}_y\right)},
\end{align}
where $\operatorname{atan2}(y,x)$ is the two-argument arctangent, which uses the signs of $x$ and $y$ separately to fix the quadrant and so returns $\Omega$ uniquely on $[0,2\pi)$, unlike $\arctan(y/x)$, whose range is the half-circle $(-\pi/2,\pi/2)$.

Let $\gamma$ be the tilt angle of the P9 velocity away from $\hat{\bm{\nu}}$, $p \equiv a_9(1-e^2)$ be the semi-latus rectum, and $u \equiv p/r = 1 +e\cos\nu$. 
We have
\begin{align}
    \label{eq:gamma_u}
    \tan\gamma & \equiv \frac{\dot{r}}{r\dot{\nu}} = -\frac{j_r}{2j_\nu} = \frac{e\sin\nu}{1+e\cos\nu},
    \\
    \label{eq:u_from_j_A_M}
    u &= \frac{j_\nu^2\,r^3}{A^2\,GM}.
\end{align}
With $e\cos\nu = u-1$ and $e\sin\nu = u\tan\gamma$,
\begin{align}
    \label{eq:inplane_elements_e}
    e &= \left[\sin^2\gamma
        + \left(\frac{u - \cos^2\gamma}{\cos\gamma}\right)^{\!2}\right]^{1/2},
    \\
    \label{eq:inplane_elements_nu}
    \nu &= \operatorname{atan2}\!\left(u\tan\gamma,\; u-1\right).
\end{align}
Then,
\begin{equation}
    a_9 = \frac{ur}{1-e^2}.
\end{equation}

The argument of perihelion follows
from the argument of latitude $\psi = \omega + \nu$, obtained by finding the counter-clockwise angle
$\hat{\bm{r}}$ makes with the ascending node $\hat{\bm{N}} \equiv \bm{R}_{\Omega, 0, 0}\hat{\bm{x}}=\cos(\Omega)\hat{\bm{x}} + \sin(\Omega) \hat{\bm{y}}$:
\begin{align}
    \label{eq:psi_atan2}
    \psi &= \mathrm{atan2}\left(\bm{\hat{r}} \cdot (\bm{\hat{l}} \times \bm{\hat{N}}) , \bm{\hat{r}} \cdot \bm{\hat{N}}\right),\\
    \label{eq:omega_from_psi_and_nu}
    \omega &= \psi - \nu.
\end{align}
Thus, we have $\bm{\theta}(\bm{\mathcal{P}})$, completing the inversion of $\bm{\mathcal{P}}(\bm{\theta})$.

\section{Acceleration-and-Jerk-only Degeneracy}
\label{sec:accel_jerk_degen}


The spin derivatives are derived from the definition of the spin phase $\phi_s$,
\begin{equation}
    \frac{d}{dt}\phi_s \equiv f_s.
\end{equation}
We have the Taylor series
\begin{align}
    \phi_s &= \phi_0 + \dot{\phi}_s t' + \tfrac{1}{2{!}}\ddot{\phi}_s t'^2 + \tfrac{1}{3 {!}}\dddot{\phi}_s t'^3 + \dots \\
    &= \phi_0 + f_s t' + \tfrac{1}{2{!}}\dot{f}_s t'^2 + \tfrac{1}{3{!}}\ddot{f}_s t'^3 + \tfrac{1}{4{!}}\dddot{f}_s t'^4 + \dots
\end{align}
where $t' = t-t_0$ is the time relative to reference time $t_0$, the time at which $f_s^{(i)}$ are measured (where $i \in \{0, 1, 2, \dots \}$).
Therefore, without detectable cubic variations in the pulse TOAs, only $\bm{a}_\odot$-based constraints via $\dot{P}_b$ measurements are possible.
With $\bm{a}_\odot$ alone, only the product $m_9/r^2$ is constrained, and $\bm{\theta}$ varies widely otherwise.
If cubic variations are significant, adding $\bm{j}_\odot$ gives six constraining equations for seven parameters, resulting in a one-parameter family of solutions.
In the absence of a positively measured $\dddot{f}_s$ and therefore the omission of $\bm{s}_\odot$, this one-parameter degeneracy remains.
Because $\ddot{f}_s$ becomes measurable well before $\dddot{f}_s$, this $\{\bm{a}_\odot, \bm{j}_\odot\}$-only regime is worth investigating. We here characterize natural choices for the remaining free parameter, each suited to different prior assumptions about the P9 parameter space.

\subsection{Knowns Across All Parameter Choices}
Before fixing a $\theta_i$ from $\bm{\theta}$ and finding the remaining $\theta_{k\neq i}$, we will explicitly state what can be uniquely determined from $\bm{a}_\odot$ and $\bm{j}_\odot$ without an additional assumption. The unit vectors $\bm{\hat{r}}$, $\bm{\hat{l}}$, and $\bm{\hat{\nu}}$ follow from \Cref{eq:basis_r,eq:basis_l}. Then, \Cref{eq:i_from_l,eq:Omega_from_l} give $i$ and $\Omega$. Thus, $\hat{\bm{N}} =\cos(\Omega)\hat{\bm{x}} + \sin(\Omega) \hat{\bm{y}}$. The angles $\omega$ and $\nu$ are not uniquely determined, but their sum $\psi=\omega + \nu$ is fixed via \Cref{eq:psi_atan2,eq:omega_from_psi_and_nu}. 
Further, while $e$ cannot be uniquely determined, it can be constrained from below using \Cref{eq:inplane_elements_e}, namely,
\begin{equation}
    \label{eq:e_constraint}
    e \geq |\sin \gamma| = \frac{|j_r|}{\sqrt{j_r^2 + 4j_\nu^2}}.
\end{equation}
The angle $\gamma$ itself is known through
\begin{equation}
    \gamma = \mathrm{atan2}(-j_r , 2j_\nu).
\end{equation}
Therefore, of $\bm{\theta}$, $\bm{\lambda} \equiv (\Omega,i, \omega+\nu)$ is known, leaving $\bm{\Lambda} \equiv (a_9, m_9, e, \{\nu \text{ or }\omega\})$. 
The following discusses treating each $\Lambda_i$ in turn as the parameter of choice. 

\subsection{The Distance or Semi-major Axis Family}

Assuming the distance $r$ from the SSB to a P9 is perhaps the most natural parameterization choice. 
Constraints on $r$ can be placed given that P9 must be faint due to its yet non-detection and given the dynamical clustering of the TNOs.
Therefore, $r$ could be sampled from these constraints in order to solve for $\bm{\Lambda}$ (see Section~\ref{sec:inverting_P_of_theta} after cubic \Cref{eq:r_cubic}). 

Assuming the semi-major axis, $\bm{\Lambda}$ is derived as follows. We have
\begin{equation}
    \label{eq:vis_viva}
    a_9 = \frac{r}{2-u \, \mathrm{sec}^2\gamma}.
\end{equation}
Equating $u$ in \Cref{eq:vis_viva,eq:u_from_j_A_M} yields
\begin{equation}
    \left(\kappa a_9 + A\right) r^3 - 2A a_9\,r^2 + GM_{\rm SS}\,r
      - 2\,GM_{\rm SS}\,a_9 = 0 ,
\end{equation}
where $\kappa \equiv (j_\nu / (A\cos\gamma))^{\!2} = \dot\nu^{\,2}\sec^2\!\gamma.$
We find this cubic has one positive root. 
With $r$, Section~\ref{sec:inverting_P_of_theta} again gives $\bm{\Lambda}$.


\subsection{The Planet Nine Mass Family}

Again, P9 must be faint and therefore cannot be too large, or, too massive. Furthermore, solar system formation and dynamical models restrict the P9 mass (e.g., \citealt{theplanetninehypothesisBB19} and references therein). Therefore, marginalizing over a distribution of masses is logical and yields a distribution in $\bm{\Lambda}$. 

As for the remainder of $\bm{\Lambda}$, we have $r = \sqrt{Gm_9/A}$, and the rest follows from above.

\begin{figure}
    \includegraphics[width=\columnwidth]{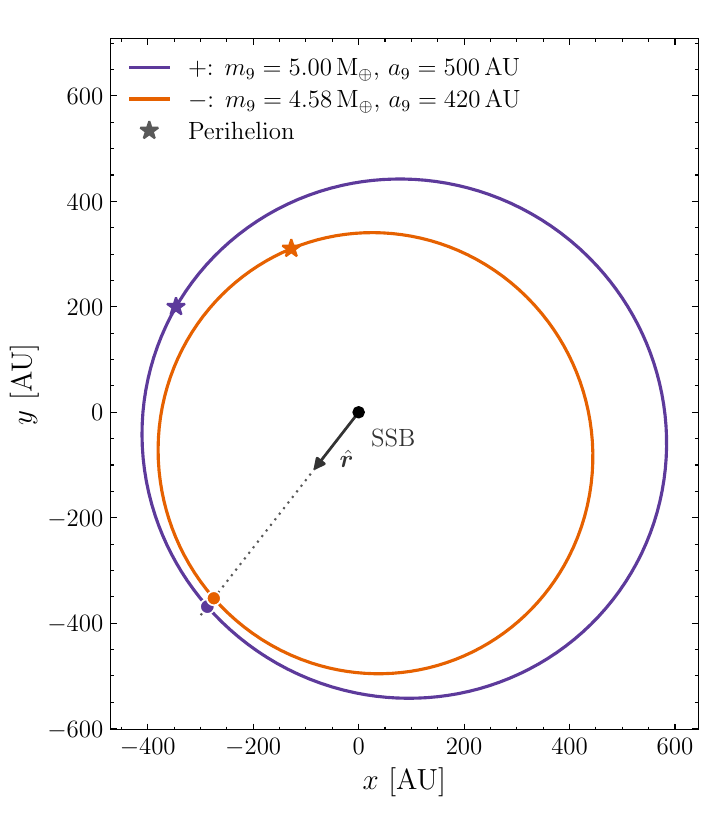}
    \caption{Two orbits that produce identical $\bm{a}_\odot$ and $\bm{j}_\odot$ at fixed $e = 0.2$. The position derivatives were calculated using $m_9 = 5\,M_\oplus$, $a_9 = 500\,\mathrm{AU}$, $i = 20^\circ$, $\Omega = 100^\circ$, $\omega = 150^\circ$, and mean anomaly of $60^\circ$. Then we solved for $\bm{\theta}$ assuming an $e=0.2$  without knowledge of $\bm{s}_\odot$, yielding two solutions, with $\bm{\theta}_+$ incidentally recovering the initial $\bm{\theta}$. For some configurations, the negative branch recovers the correct solution given a correct $e$ assumption.}
    \label{fig:branches}
\end{figure}

\subsection{The Two Eccentricity Families}

The eccentricity determines the shape of the orbit and how long we expect a P9 to spend near its aphelion. However, fixing an eccentricity does not constrain a single unique orbit, but rather two potential orbits. 

From the definition of $\gamma$ in \Cref{eq:gamma_u} and $u = 1 + e \cos \nu$, we have
\begin{equation}
    \label{eq:u_parabola}
    u^2\sec^2\gamma - 2u + \left(1-e^2\right) = 0,
\end{equation}
which we can solve for $u$ given that each coefficient is known. We have
\begin{equation}
    u_\pm = \cos^2\gamma \pm \cos\gamma\sqrt{e^2 - \sin^2\gamma}.
\end{equation}
Recall that $e \geq | \sin \gamma |$ and so this has one real solution for $e=0$ and two solutions otherwise.
\Cref{eq:u_from_j_A_M} gives $r_\pm$, which gives $\bm{\Lambda}_\pm$. Thus, one cannot marginalize over $e$ alone, but must also either include another element of $\bm{\Lambda}$ or choose a branch of $u_\pm$. We illustrate an example of this non-uniqueness in Figure~\ref{fig:branches}.


\subsection{The True Anomaly or Argument of Perihelion Families}

For eccentric orbits, $\nu$ spends more time near $180^\circ$ than $0^\circ$. Furthermore, clustering of the apsidal lines of TNOs paired with the fact that P9 is suspected to differ in $\omega$ by $180^\circ$ from its potentially sculpted TNOs \citep{theplanetninehypothesisBB19} gives reason to assume a value for $\omega$. 

We have 
\begin{equation}
    e = \frac{\sin \gamma}{\sin{(\nu - \gamma)}}
\end{equation}
and then $u = 1 + e \cos \nu$. The rest of $\bm{\Lambda}$ follows from the previous subsection. 

\section{Estimated Constraint Space}
\label{sec:estimated_constraint_space}

We will now discuss what preliminary constraints current pulsar data sets can place on P9 using the formalism proposed in Section~\ref{sec:methods}. 
We here use $\dot{P}_b$ measurements from \citeauthor{donlanv3} (\citeyear{donlanv3}; D+25 hereafter)\footnote{This work used version 3 at \url{https://github.com/thomasdonlon/Empirical_Model_MSP_Spindown_Accels}} and $\ddot{f_s}$ measurements from L+19. 
Let $\mu_{a}^{\beta} = \dot{P}_b^{9, \beta} / P_b^\beta$ and $\mu_{j}^\beta = \ddot{f}^\beta_s / f^\beta_s$ for pulsar $\beta$, and $\sigma_\alpha^\beta$ be the uncertainty in measuring $\mu_\alpha^\beta$ for $\alpha \in \{a, j\}$. Note that relatively few pulsars in D+25 and L+19 have both $\dot{P}_b$ and $\ddot{f}$ measurements.
Using the circular orbit ($e=0$) forms of \Cref{eq:accel_ssb,eq:jerk_ssb}, we have the best-case $\mu_\alpha^\beta$ for each pulsar
\begin{align}
    \mu_a^{\mathrm{max}} & \approx \frac{Gm_9}{c a_9^2},\\
    \mu_j^{\mathrm{max}} & \approx \sqrt{GM}\frac{Gm_9}{c a_9^{7/2}}.
\end{align}
For reference, 
\begin{align}
    \nonumber
    \frac{\mu_a^{\mathrm{max}}}{8.9\times 10^{-20}\, \mathrm{Hz}} &\approx  \left(\frac{m_9}{15\,\mathrm{M}_\oplus} \right) \left( \frac{100\, \mathrm{AU}}{a_9} \right)^{2} \\
    &\approx \left( \frac{\dot{P}_b}{7.7\times 10^{-15} \, \mathrm{s/s}} \right) \left( \frac{1 \, \mathrm{d}}{P_b} \right),\\
    \nonumber
    \frac{\mu_j^{\mathrm{max}}}{1.8\times 10^{-29}\, \mathrm{Hz^2}} &\approx  \left(\frac{m_9}{15\,\mathrm{M}_\oplus} \right) \left( \frac{100\, \mathrm{AU}}{a_9} \right)^{7/2} \\
    &\approx \left( \frac{\ddot{f}_s}{3.5\times 10^{-27} \, \mathrm{Hz}^3} \right) \left( \frac{200 \, \mathrm{Hz}}{f_s} \right).
\end{align}
That is, a P9 with $m_9=15\,\mathrm{M}_\oplus$ and $a_9=100$\,AU would induce a $\ddot{f}_s \approx 3.5\times10^{-27}$\,Hz$^3$ on a $f_s = 200$\,Hz pulsar when the pulsar and orbital geometries are most favorable.
The best-case signal-to-noise ratio (SNR) of pulsar $\beta$ and positional derivative term $\alpha$ is \begin{equation}
    \mathcal{S}_\alpha^\beta \equiv \frac{\mu_\alpha^{\mathrm{max}}}{\sigma _\alpha^\beta}.
\end{equation}
The overall SNR for channel $\alpha$ is 
\begin{equation}
    \mathcal{S}_\alpha = \left[ \sum_\beta \left( \mathcal{S}_\alpha^\beta \right)^2\right]^{1/2},
\end{equation}
and the total SNR across both position terms and all pulsars is 
\begin{equation}
    \mathcal{S}_\mathrm{tot} = \left[ \sum_{\alpha,\beta} \left( \mathcal{S}_\alpha^\beta \right)^2\right]^{1/2} = \sqrt{(\mathcal{S}_a)^2 + (\mathcal{S}_j)^2}.
\end{equation}

\begin{figure*}
  \centering
  \includegraphics[width=0.95\textwidth]{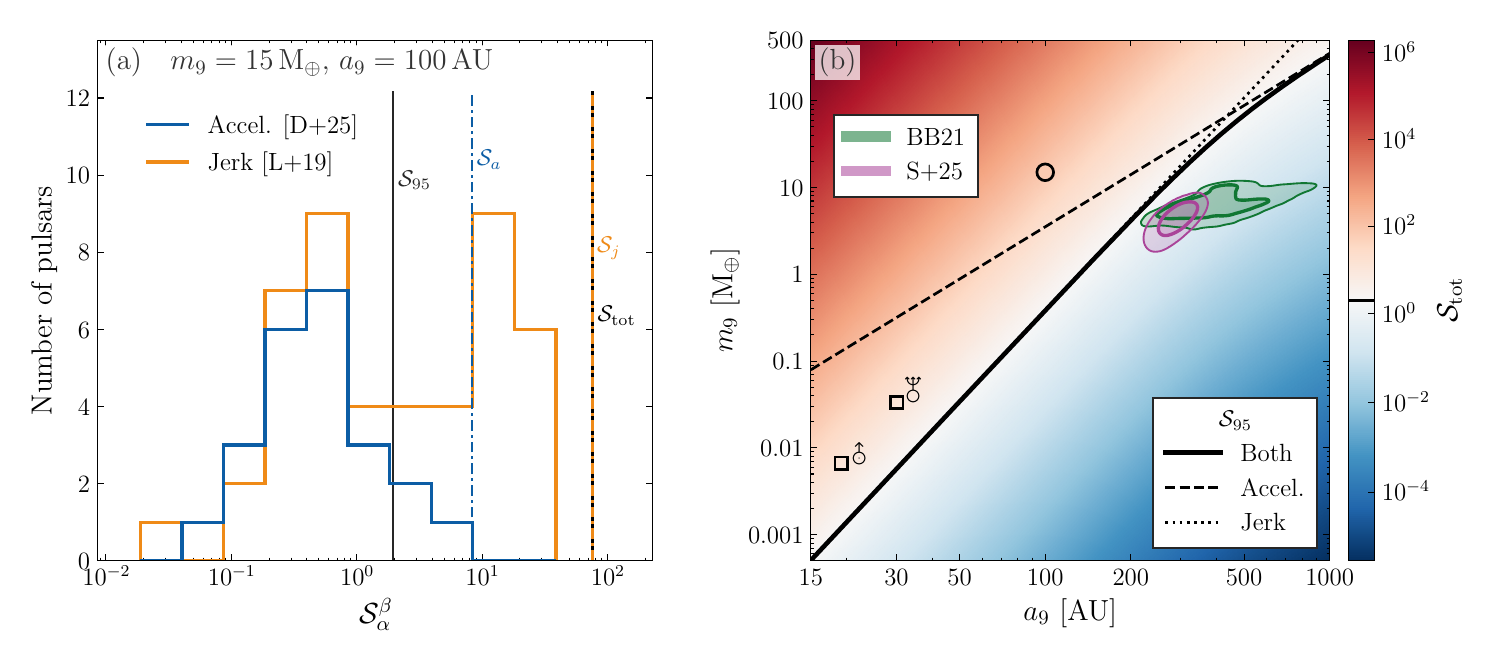}
  \caption{SNRs of the two position channels, for a circular orbit and the most favorable line-of-sight geometry. 
  $\dot{P}_b$ and $\ddot{f}_s$ measurements come from \citet{donlanv3} and L+19, respectively.
  (a) Per-pulsar $\mathcal{S}_\alpha^\beta$ at $m_9=15\,\mathrm{M}_\oplus$ and $a_9=100$\,AU. The four vertical lines mark different SNRs as defined in the text.
  (b) $\mathcal{S}_{\rm tot}$ over the $(a_9,m_9)$ plane; the contours mark where each channel individually, and the two combined, reach $\mathcal{S}_{95}$. 
  Overlaid are the 68\% and 95\% credible regions of \citet{brown2021orbit_BB21} (BB21) and \citet{siraj2025orbit} (S+25), marginalized onto the $(a_9, m_9)$ plane.
  The hollow circle at (100\,AU, 15\,M$_\oplus)$ is the point used in panel (a). Pulsar-based mass constraints on known solar system bodies also follow this curve (Section~\ref{sec:uranus_neptune}). The 95\% mass constraint ranges on Uranus (\uranus) and Neptune (\neptune) found in Section~\ref{sec:uranus_neptune} are denoted by the hollow squares, with ranges above the theoretical best $\mathcal{S}_\mathrm{tot} = \mathcal{S}_{95}$ as expected. Additional pulsars and longer baselines are expected to improve $\mathcal{S}_\mathrm{tot}$.}
  \label{fig:constraint_space}
\end{figure*}

An SNR above
$\mathcal{S}_{95} \approx1.96$
marks when the available data can \emph{just} begin to make constraints on the P9 parameter space ($\bm{\theta}$) at the 95\% confidence level under an optimal pulsar sky distribution for a given P9. 
Looking at $\sigma_\alpha^\beta$ from D+25 and L+19, and ignoring asymmetries in $\mu_\alpha^\beta \pm \sigma_\alpha^\beta$ around zero (i.e., ignoring $\mu_\alpha^\beta \neq 0$), we plot their SNRs in Figure~\ref{fig:constraint_space}. The left panel shows a histogram of $\mathcal{S}_\alpha^\beta$ for a fixed $(a_9,m_9) = (100\,\mathrm{AU}, 15\,\mathrm{M_\oplus})$. We note that the histograms of $\sigma_a^\beta$ and $\sigma^\beta_j$ would look proportionally the same internally as $\mathcal{S}_a$ and $\mathcal{S}_j$, respectively, but with different units. The right panel plots $\mathcal{S}_\mathrm{tot}$ over a portion of the $(a_9,m_9)$ plane, with contours of $\mathcal{S}_a = \mathcal{S}_j = \mathcal{S}_\mathrm{tot} = \mathcal{S}_\mathrm{95}$ shown. 
We find that $\mathcal{S}_a = \mathcal{S}_j$ at $a_9 \approx 450$\,AU at any mass for the dataset used. We overlay the expected P9 parameter space marginalized onto the $(a_9,m_9)$ plane as found by BB21 and \citet{siraj2025orbit} (S+25 hereafter). We find that $\mathcal{S}_\mathrm{tot}$ needs to improve by factors of roughly ($1.3,1.7$) and ($1.6,2.2$) to begin to inform the ($95\%, 68\%$) enclosed spaces of BB21 and S+25, respectively.

Based on how $\mathcal{S}_{\mathrm{tot}}$ is defined and constructed, we do not expect that $\bm{\theta}$ can be well constrained where its $(a_9,m_9)$ projection sits below the $\mathcal{S}_\mathrm{95}$ curve in Figure~\ref{fig:constraint_space}. Further, the $\bm{\theta}$ space sitting above but close to the $\mathcal{S}_{\mathrm{95}}$ curve may not actually be constrained to a 95\% significance for any random, finitely-numbered pulsar sky distribution. This is particularly true for a realistic known-MSP distribution, which is denser in the Galactic plane and towards the Galactic center. Therefore, there is a complex position and orbital-plane dependence of a P9 (see \Cref{eq:accel_ssb,eq:jerk_ssb} and the sky positions in Table~1 of L+19), the nature of which we save for a forthcoming paper. An analysis of more-recent, longer-baseline data and a larger array of pulsars will also be presented in a forthcoming paper.


\section{Weighing Uranus, Neptune, and a Future Candidate Planet Nine}
\label{sec:uranus_neptune}

Upon a future, potential optical detection of a P9, the orbital parameters, barring a model-dependent mass, could be constrained within a few years, or sooner if the P9 is found in archival data. 
Given these orbital parameters, the formalism presented in Sections~\ref{sec:methods} and \ref{sec:accel_jerk_degen} can also be used to constrain the mass of any planet in the solar system.
Here, we present a method for weighing the ice giant systems of Uranus and Neptune from their known orbital parameters, an approach that would translate immediately to a newly detected P9.


Figure~\ref{fig:constraint_space} includes part of the ($m_9 < 1$\,M$_\oplus$, $a_9 < 100$\,AU) space to showcase how our method could be used to constrain the residual masses of Uranus ($a_7=19.2$\,AU) and Neptune ($a_8=30.1$\,AU), after swapping $m_9$ with $\delta m_7$ and $\delta m_8$, where $\delta m_k$ is the residual mass from the assumed $m_k$ of the used ephemeris. \Cref{eq:accel_ssb,eq:jerk_ssb} still hold to leading order since $m_0=\mathrm{M}_\odot \gg m_{k\neq0}$, and M$_\mathrm{SS}$ can be modified to exclude $m_7$ or $m_8$. Pulsar timing has already yielded independent estimates on the mass of the major solar system bodies (e.g., \citealt{champion2010measuring}; C+18). Therefore, we present the following as a consistency check on our own method. 

Let $\bm{\theta}_k'$ be $\bm{\theta}_k$ without $m_k$, where $\bm{\theta}_k$ is the $\bm{\theta}$ of object $k$. $\bm{\theta}'_k$ is known from the solar system ephemerides (we choose, as do L+19, the DE421 planetary ephemeris; \citealt{de421}).
At the orbital distance of Neptune and closer, the jerk measurements alone contribute to the SNR (see Figure~\ref{fig:constraint_space}). Therefore, we fix $\alpha = j$ and ignore the $\dot{P}_b$ dataset. The likelihood, as a function of $\bm{\delta m} \equiv(\delta m_7, \delta m_8)$, is
\begin{align}
    \label{eq:jerk_likelihood}
    \mathcal{L}_{j, \{k=7,8\}}^\beta & \propto \exp\left\{ -\frac{1}{2}
      \left[ \frac{\left(\bm{j}_{\odot,7} + \bm{j}_{\odot,8}\right)
             \cdot\hat{\bm{n}}^\beta/c - \mu_j^\beta}
            {\sigma_j^\beta} \right]^{2} \right\},\\
    \label{eq:final_likelihood}
    \mathcal{L} \left(\bm{\delta m}\right) &= \prod_{\beta} \mathcal{L}_{j, \{k=7,8\}}^\beta,
\end{align}
where 
\begin{equation}
    \bm{j}_{\odot,k} = \bm{j}_\odot(\delta m_k, \bm{\theta}'_k)
\end{equation}
as given by \Cref{eq:jerk_ssb}.
$\delta m_k$, not $m_k$, enters into $\bm{j}_\odot$ as the best-estimate mass is already in the ephemeris used to measure $\ddot{f}_s$. We are measuring residual deviations from this mass, which would induce a dipolar jerk signal. 
Setting a uniform prior (i.e., $P(\bm{\delta m}) \propto 1$) yields the posterior
\begin{equation}
    \label{eq:posterior}
    P\left(\bm{\delta m} | \bm{d} \right) \propto \mathcal{L} \left(\bm{\delta m}\right),
\end{equation}
where $\bm{d}$ is the set of all $\mu_j^\beta$ and $\sigma_j^\beta$. To include $\dot{P}_b$ data, we would construct a $\mathcal{L}_{a,k}^\beta$ analogous to \Cref{eq:jerk_likelihood} and include it in the \Cref{eq:final_likelihood} product. 

\begin{figure}
    \includegraphics[width=\columnwidth]{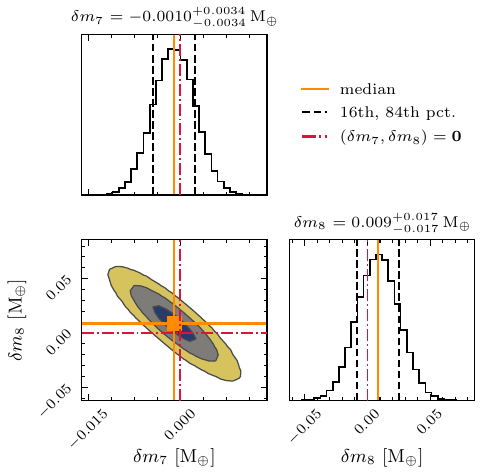}
    \caption{Joint $(\delta m_7, \delta m_8)$ posterior as defined by \Cref{eq:posterior}. Results are consistent with zero (dot-dashed lines).} 
    \label{fig:uranus_neptune_posterior}
\end{figure}

We sample the posterior \Cref{eq:posterior} using Markov-Chain Monte Carlo methods and present our results in Figure~\ref{fig:uranus_neptune_posterior}. 
We use $\bm{\theta}'_k$ information from DE421 at the reference epoch of each pulsar's $\ddot{f}_s$ measurement in L+19.
There is a large covariance between the two mass measurements, due to their relatively close sky position currently and their near-sharing of the same orbital plane.
Marginalizing one mass measurement over the other, we find
$G(\delta m_7)/(\mathcal{GM})^{\rm N}_{\rm E} = -0.001(3)$ and $G(\delta m_8)/(\mathcal{GM})^{\rm N}_{\rm E} = 0.009(20)$. 
These uncertainties (multiplied by $\mathcal{S}_{95}$) lie above the $\mathcal{S}_\mathrm{tot} = \mathcal{S}_{95}$ curve for the assumed positions and orbital planes of Uranus and Neptune and the selected pulsar sample. 
The total masses in solar units, after adding the reference masses from DE421, are tabulated in Table~\ref{tab:icegiant_masses}.
$\bm{\delta m}$ is consistent with zero and the pulsar-based mass measurements of C+18. Indeed, our precision is nearly identical to the work of C+18, with slightly different medians. 
The consistency of $\boldsymbol{\delta m}$ with zero, at a precision comparable to that of C+18, also suggests that the L+19 $\ddot{f}_s$ uncertainties are not severely underestimated.
Therefore, our method should prove to give competitive future constraints on the known solar system bodies with relatively light $\ddot{f}_s$ datasets alone, rather than requiring large TOA datasets.

\begin{table}
    \centering
    \caption{Uranus and Neptune system masses from pulsar timing, as ratios to the nominal solar gravitational mass $(\mathcal{GM})^{\rm N}_\odot$. Our results are in agreement with \citet{caballero2018studying}, denoted by C+18.}
    \label{tab:icegiant_masses}
    \begin{tabular}{lcc}
        \hline
        & \multicolumn{2}{c}{$(Gm_k)/(\mathcal{GM})^{\rm N}_\odot$ $\left[10^{-5}\right]$} \\
        \cline{2-3}
        & Uranus & Neptune \\
        \hline
        C+18, DE421 & 4.368(1) & 5.144(6) \\
        This work, DE421             & 4.366(1) & 5.154(5) \\
        \hline
    \end{tabular}
\end{table}

An optical detection of P9 will provide $\bm{\theta}'_9$. 
Uncertainties in $\bm{\theta}'_9$ can be accounted for via sampling and marginalizing over a hyperprior (see, e.g., \citealt{valtolina2025regularizing_hyperparamaters}). 
We ignored uncertainties in $\bm{\theta}'_k$ in the case of Uranus and Neptune as their orbits are far better constrained than $\ddot{f}_s$ measurements. 
The likelihood $\mathcal{L}_{\alpha,9}^\beta (m_9)$ constructed similarly to \Cref{eq:jerk_likelihood} can constrain $m_9$ independent of the P9 albedo, radius, and thermal emission.

\section{Discussion and Conclusions}
\label{sec:discussion_and_conclusions}

We have presented a method to search for a distant Planet Nine. We show that matching the P9-induced SSB position derivatives to Doppler terms of increasing order allows for a unique determination of the P9 parameter space $\bm{\theta}$. 
We invert this relation uniquely.
Therefore, because each pulsar contributes only one or two derived quantities, our method operates on light datasets of $\mathcal{O}(10^2)$ entries in total, rather than the $\mathcal{O}(10^4)$ TOAs per pulsar in a full timing analysis.
By weighing the ice giant systems, we show that our approach provides the same competitive constraints as those from previous studies. 
Further, we found analytic expressions for the degeneracies that arise when only some Doppler terms are measurable.


Showing that cubic and quartic variations in pulse TOAs are required to determine an orbit is a key result from this study. 
The full Kepler equations of motion or N-body integrations provide more accurate results over longer timing baselines, but omit the intuitive understanding for why certain parameters will be poorly constrained or dominated by the prior probability belief. 
Our work describes how each positional derivative informs the orbital parameters. The acceleration informs the current SSB-centered direction of P9, as well as $m_9/r^2$. Adding jerk defines the orbital plane, gives a lower bound on the eccentricity, and leaves a one-parameter family of orbits. Lastly, including snap closes the degeneracy by measuring $M/r^3$, while providing two over-constraining equations. If the snap is found to be out-of-plane from $\bm{a}_\odot \times \bm{j}_\odot$, the P9 model would not explain the positional terms, therefore questioning the model in deriving $\bm{\theta}(\bm{\mathcal{P}})$. Other models would have to be considered, such as multiple planet-mass TNOs, mismodeled known solar system objects, or mismodeled spin frequency derivatives.


The $\{\bm{a}_\odot, \bm{j}_\odot\}$-only regime is not merely an academic concern. $\ddot{f}_s$ in MSPs becomes measurable long before $\dddot{f}_s$, meaning that omission of snap, or, quartic TOA variations, leaves a unique pulsar-derived P9 parameter space for a future dataset and analysis. However, there are still useful insights from $\{\bm{a}_\odot, \bm{j}_\odot\}$ that we briefly discussed in Section~\ref{sec:accel_jerk_degen}. For starters, while the eccentricity family can be reasonably selected, it must be understood that two orbits, not one, are specified. Choosing priors on the distance, semi-major axis, true anomaly, and/or the argument of perihelion are also possible and reasonable, and can be informed by the P9 space constrained through other lines of evidence. 


The Vera C. Rubin Observatory may discover the P9 as currently conceived by \citet{theplanetninehypothesisBB19} by the end of the decade. While the physical location of this P9 would be known through direct observations, unless P9 hosts a natural satellite, mass constraints will depend on features intrinsic to the planet, such as albedo, shape, thermal emission, and potentially its nature as a primordial black hole as suggested by \citet{primordial_black_hole_2020}. Therefore, in Section~\ref{sec:uranus_neptune} we outline a method to weigh P9 by fixing the physical orbital parameters $\bm{\theta}'_9$. This would also escape the under-constrained $\{\bm{a}_\odot, \bm{j}_\odot\}$-only problem, so properly modeled accelerations or jerks would be sufficient.


Whether a P9 is discovered in the next decade or not, there is more work to be done in terms of pulsar-based constraining power. Figure~\ref{fig:constraint_space} shows that a $\sim$2019 dataset featuring 87 pulsars is already on the cusp of informing the expected P9 parameter space. The exact timeline will depend on the sample of pulsars analyzed and their timing baselines. An analysis using the North American Nanohertz Gravitational Wave Observatory 15\,yr dataset \citep{agazie2023nanograv} or upcoming 20\,yr dataset would roughly double the pulsar sample size, and push the baseline up roughly four years from L+19. Using data from the upcoming International Pulsar Timing Array DR3, rather than the DR2 of \citet{IPTA_DR2_2019}, would improve the sensitivity even further. 

Because pulsar timing and TNO dynamics constrain P9 through independent data, their likelihoods can be combined directly. Once pulsar timing constrains the P9 parameter space at a level comparable to TNO dynamical analyses (e.g., BB21; S+25), a joint pulsar–TNO analysis will yield the tightest constraints on P9 yet. 
The coming decade should be an exciting time for the study of the solar system.

\begin{acknowledgments}
We thank T. Joseph W. Lazio, Matthew Cerep, and Jordan O'Kelley for helpful discussions. 
Computational resources were provided by the WVU Research Computing Harpers Ferry HPC cluster, which is funded in part through a NASA Congressionally Directed Spending.
J.D.T. and E.F. are supported by the National Science Foundation under grant PHY-2607948.
The authors drafted the initial manuscript text and figures, and then  made limited use of Claude Opus 5.5 (Anthropic) to suggest wording and figure code improvements. All text, code, and figures were reviewed and revised by the authors, who take sole responsibility for this work.
\end{acknowledgments}

\begin{contribution}

J.D.T. conceived of and led the project, as well as produced the figures, prepared the manuscript, and conducted the analysis, with input from E.F. and J.G. E.F. and J.G. contributed insightful discussions. J.G. provided computational support. 


\end{contribution}

%


\software{\texttt{astropy} (Astropy Collaboration et al. \citeyear{astropy:2013}, \citeyear{astropy:2018}, \citeyear{astropy:2022}),
\texttt{PINT} \citep{luo2019ascl, luo2021pint},
\texttt{numpy} \citep{HarrisMillman+2020}, 
\texttt{matplotlib} \citep{Hunter2007}, 
\texttt{corner} \citep{Foreman-Mackey2016}.}





\bibliography{main}{}
\bibliographystyle{aasjournalv7}



\end{document}